# Modified EESM Based Link Adaptation Algorithm for Multimedia Transmission in Multicarrier Systems

R. SANDANALAKSHMI[1], ATHILAKSHMI[2], K. MANIVANNAN[3]

[1,2]Department of Electronics and Communication Engineering
Pondicherry Engineering College
Pondicherry, India

[3]Department of Electrical & Electronics Engineering
Pondicherry Engineering College
Pondicherry, India

**Abstract**

The previous link adaptation algorithms on ofdm based systems use equal modulation order for all sub carrier index within a block. For multimedia transmission using ofdm as the modulation technique, unequal constellation is used within one ofdm subcarrier block, a set of subcarriers for audio and another set for video transmissions. A generic model has been shown for such a transmission and link adaptation algorithm has been proposed using EESM (Effective Exponential SNR mapping) method as basic method. Mathematical model has been derived for the channel based on bivariate Gaussian distribution in which the amplitude varies two dimensionally in the same envelope. From the Moment generating function of bivariate distribution, Probability of error has been theoretically derived. Results have been shown for BER performance of an ofdm system using unequal constellation. BER performances have been shown for different values of correlation parameter and fading figure.

## 1. Introduction

Next generation cellular systems support multiple transmission modes, which can be used to improve the performance of such systems by adapting to current channel conditions. This process is referred to as link adaptation. Typically, these transmission modes include different modulation and coding, schemes (MCS) and different multiple antenna arrangements modes – such as beam-forming, space- time coding and spatial multiplexing – as the transmission becomes multidimensional in space, time, and frequency domain. Orthogonal Frequency Division Multiplexing (OFDM) is the air interface for 802.11, 802.16 (WiMAX), and 3GPP Long Term Evaluation (LTE) systems. The resources typically referred to as subcarriers, available in an OFDM frame, can be defined on a time frequency grid [1]. The performance of a binary code depends on the channel condition obtained over the allocated subcarriers. Typically, the channel is frequency selective, and a mean SNR metric is only sufficient to obtain a long term performance metric of the channel. On the other hand, short term performance metrics, which are also key to obtaining performance enhancements with feedback, are obtained from the actual instantaneous channel realization.

A well-known approach to link performance modeling and link quality prediction is the Effective Exponential SINR Metric (EESM) method, which computes an effective SNR (also referred to as AWGN equivalent SNR) metric by taking as input the individual sub carrier SNRs and using an exponential combining function. Once computed, the block error rate is obtained from looking up an AWGN performance curve. This approach has been widely applied to OFDM link layers and is based on the performance approximation by asymptotic union bounds [2]. It has been shown in literature survey [5, 6] that the Doppler spreading of OFDM systems follows joint probability distribution function implies the transmission will also be dependent on Bivariate Gaussian distribution and gives better PDF, CDF. Under the same assumption the amplitude variations of the channel under two different constellations has been modeled as bivariate Gaussian distribution. The paper has been organized as follows; Chapter 1 explains the link adaptation based on unequal constellation. Chapter II derives the mathematical model based pn bivariate Gaussian distribution, and Chapter III produces the simulated results.

## 2. EESM for Unequal Constellation in OFDM Block

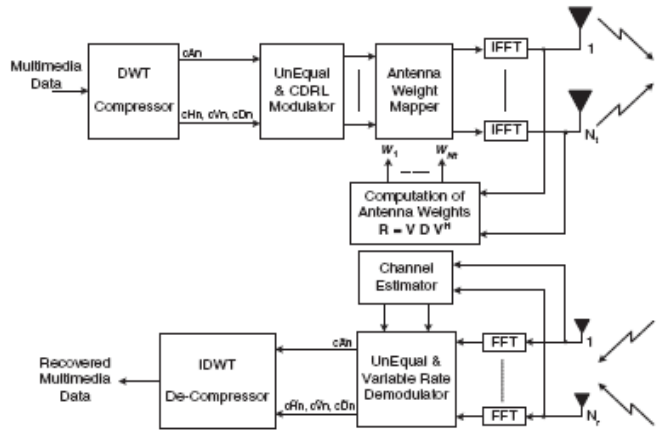

Fig1.Unequal ofdm transceiver structure

The proposed structure [15] transceiver structure consists of data compressor, unequal adaptive modulator and channel estimator. Multimedia data is compressed using a source coder based on discrete wavelet transformation, in wavelet analysis we represent low frequency information by approximation coefficients cAn an high pass spatial frequency data by horizontal vertical an diagonal signal by cHn,cVn,cDn. To achieve a target SNR we allocate approximate coefficients with low modulation order and detail coefficients with high modulation order based on the channel estimation feedback message from the receiver.

EESM method has been identified as one of the fast link adaptation technique for multicarrier based systems. For a fast link adaptation of multimedia transmission where the audio and video are transmitted in different constellations on the same ofdm block, the EESM method has been modified and used for performance prediction (e.g., the FER metric) for the current channel conditions. In the subcarrier block with two modulation order are taken such a way that low frequency components (Audio) are transmitted with low modulation order and high frequency components (video) with high modulation order. This method maps a set of per subcarrier SNRs { $\gamma_1 \ldots \gamma_{N2}$} to a single effective SNR (SNReff) :

$$\gamma_{eff} = -\beta . \ln \left[ \frac{\sum_{i=1}^{N1} e^{-\frac{\gamma_i}{\beta 1}} + \sum_{i=N1+1}^{N2} e^{-\frac{\lambda_i}{\beta 2}}}{N1+N2} \right] \quad (1)$$

Where N1&N2 is the number of sub-carriers used by a codeword , for 1 to N1 subcarriers β1 is the calibration parameter and from N1+1 to N2 subcarriers β2 is used as the calibration parameter that is typically different for each MCS. The procedure used to link adaptation with EESM is listed herewith

Determine quadratic approximation of SINR Vs β.
1. Obtain the parameters a,b,c from quadratic equation
    SNR$_{eff}$ (β) = a + b β$_{db}$ + c β$^2_{db}$
    Where a,b,c are y-intercept ,linear and quadratic parameters.
2. Sent a,b,c as Channel quality indicator to BTS.
3. BTS receiving the parameters will update it and select the appropriate β for the required SNReff .
4. Based on current β value the modulation order is selected.

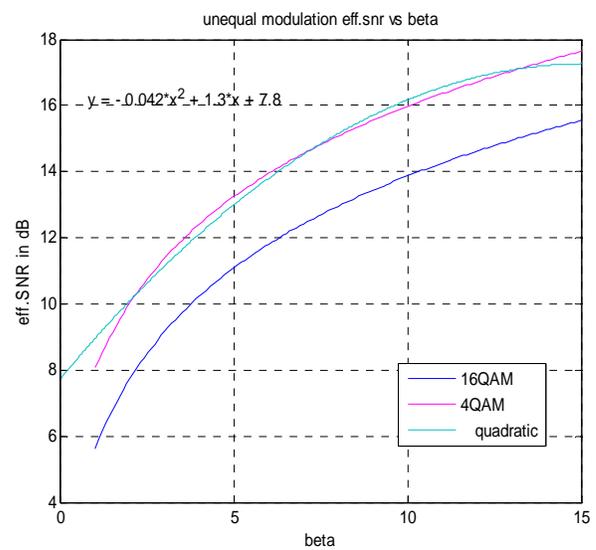

Fig. 2. Instantaneous SNReff vs. β1 and SNReff vs. β2 curve with the corresponding quadratic approximations.

## 3. Vertical Shift Method

Enabling the method described above it would require the approximation of the SNReff vs. β1 and curve to SNReff vs. β$_2$ have to be sent potentially as often as every frame, in order to track changes in the SNR due to fading. This represents less feedback than sending the entire channel response {γ 1, …, γN}, but is still a significant amount of
                                …. (1)

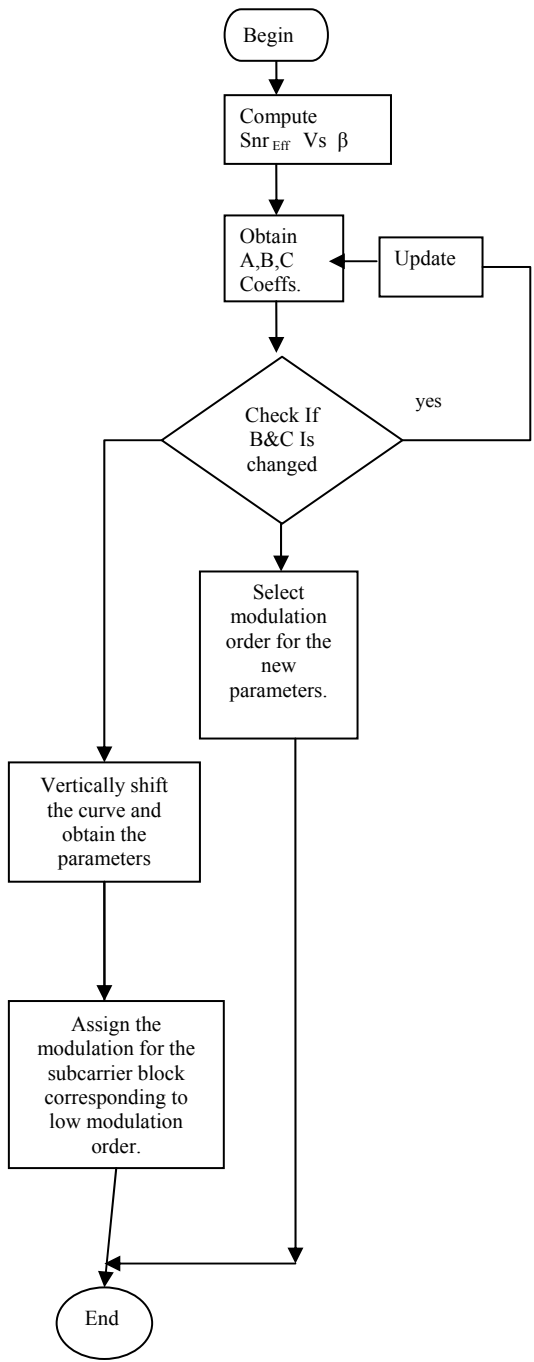

Figure 3. Modified EESM link adaptation technique for unequal constellations

information. For the quadratic approximation, three coefficients (a, b, c) for two different modulation orders would have to be fed back every frame. A first possible approach to eliminate the frame-by-frame feedback of curve parameters is to assume that the shape of the β curve does not change significantly with changes in band-average SNR as long as the channel power-delay profile does not change.

More specifically, this method consists of sending infrequently the instantaneous SNReff vs. β1 curve only as a reference SNR value (reference curve) instead of transmitting two curves. The SNReff vs. β2 curve for the current channel is then obtained by a simple vertical (1-dimensional) shift of the reference curve by 3dB. This first order approximation will be referred to as the vertical shift method.

## 3. Unequal Modulations – Bivariate Distribution

The two dimensional amplitude variations within same channel envelope can be modeled as bivariate Gaussian distribution. The purpose of such a multilevel modulation is that the carriers in an ofdm block which can withstand deep fading can be allotted a low modulation order and the others high level modulation and as the result the SNR performance can be improved much better than equal modulation. From theory two univariate marginal distributions following Gaussian distribution can be modeled as under the same roof as bivariate joint distribution given [3] by

$$F = \frac{(x_1 x_2)^{(\alpha-1)/2}}{\Gamma(\alpha)(\beta_1 \beta_2)^{(\alpha+1)/2}(1-\rho)\rho^{(\alpha-1)/2}} \times \exp\left(-\frac{x_1/\beta_1 + x_2/\beta_2}{1-\rho}\right) \times I_{\alpha-1}\left(\frac{2\sqrt{\rho}}{1-\rho}\sqrt{\frac{x_1 x_2}{\beta_1 \beta_2}}\right) U(x_1) U(x_2)$$

(2)

Where Γ(.) is gamma function and I(.) is a modified Bessel function of first kind ,β1 and β1 are scaling parameters, x1 and x2 are random variables for different marginal distributions, and ρ is the correlation coefficient ,When ρ → 0 the joint pdf tends to product of two univariate gamma distribution . Consider two OFDM modulation techniques 16QAM and 64QAM and the joint probability distribution for them is given below,

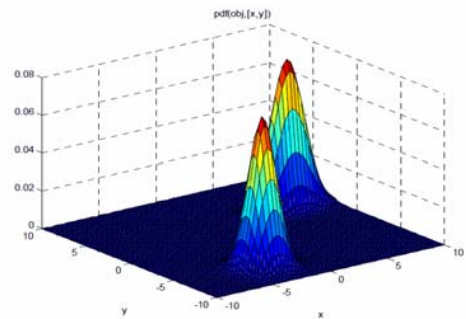

Fig. (4) PDF for Bivariate normal distribution for 16QAM and 64QAM

Since the channel is characterized by two different constellations, the channel has two different parameterized fading envelope. Hence, the envelope of the channel response can be modeled as nakagami-m model which is characterized by two parameters, fading figure and moments. It has the inherent advantage of versatility and covers a wide range of fading channel scenarios. The m-distributed pdf of the envelope is taken as [2].

$$P(R) = \frac{2m^m R^{2m-1}}{\Gamma(m)\Omega^m} \exp(-\frac{mR^2}{\Omega}), R \geq 0$$

$$\text{where } \Omega = E(R^2), m = \frac{\Omega^2}{E[(R^2-\Omega)^2]}, m \geq 1/2$$

(3)

E( ) represents the average and $\Gamma$( ) represents the gamma function, $\Omega/2$ is the average power of the signal, 'm' is named fading figure, parameter related to fading range. In the case of M-distributed correlated nonidentical fading Nakagami-$m$ channels PDF of the combined signal envelope[14]- $pa(rt)$, is given by

$$p_a(r_t) = \frac{2r_t \sqrt{\pi}}{\Gamma(m)[\sigma_1\sigma_2(1-\rho)]^m} \left[\frac{r_t^2}{2\beta}\right]^{m-1/2}$$

$$I_{m-1/2}(\beta r_t^2) e^{-\alpha r_t^2}, r_t \geq 0 \quad (4)$$

where $I\upsilon(\cdot)$ denotes the $\upsilon$th-order modified Bessel function

$$\rho = \frac{\text{cov}(r_1^2, r_2^2)}{\sqrt{\text{var}(r_1^2)\text{var}(r_2^2)}}, 0 \leq \rho < 1 \quad (5)$$

is the envelope correlation coefficient [3] between the two signals[6] and the parameters $\sigma_d$ ($d = 1, 2$), $\alpha$, and $\beta$ are defined as follows:

$$\sigma_d = \frac{\Omega_d}{m}, d = 1, 2$$

$$\alpha = \frac{\sigma_1 + \sigma_2}{2\sigma_1\sigma_2(1-\rho)}$$

$$\beta^2 = \frac{(\sigma_1 - \sigma_2)^2 + 4\sigma_1\sigma_2\rho}{4\sigma_1^2\sigma_2^2(1-\rho)^2}$$

(6)

where $\Omega_d$, $d = 1, 2$, is the average fading power of the uneven sub-carrier blocks. The PDF of the combined SNR per symbol, $p_a(\gamma_1)$ is given by

$$p_a(\gamma_1) = \frac{\sqrt{\pi}}{\Gamma(m)} \left[\frac{m^2}{\bar{\gamma}_1\bar{\gamma}_2(1-\rho)}\right]^m \left(\frac{\gamma_1}{2\beta'}\right)^{m-1/2} I_{m-1/2}(\beta'\gamma_1) e^{-\alpha'\gamma_1}$$

$$\gamma_1 \geq 0 \qquad (6)$$

Where the parameters $\alpha'$ and $\beta'$ are normalized versions of the parameters $\alpha$ and $\beta$, and are given by

$$\alpha' = \frac{\alpha}{E_s/N_0} = \frac{m(\bar{\gamma}_1 + \bar{\gamma}_2)}{2\bar{\gamma}_1\bar{\gamma}_2(1-\rho)}$$

$$\beta' = \frac{\beta}{E_s/N_0} = \frac{m\left((\bar{\gamma}_1 + \bar{\gamma}_2)^2 - 4\bar{\gamma}_1\bar{\gamma}_2(1-\rho)\right)^{1/2}}{2\bar{\gamma}_1\bar{\gamma}_2(1-\rho)}$$

(7)

and finally the equation reduces to

$$p_a(\gamma_1) = \frac{1}{2\sqrt{\rho}\bar{\gamma}} \left\{\exp\left[-\frac{\gamma_1}{(1+\sqrt{\rho})\bar{\gamma}}\right] - \exp\left[-\frac{\gamma_1}{(1-\sqrt{\rho})\bar{\gamma}}\right]\right\},$$

$$\gamma_1 \geq 0 \qquad (8)$$

Using the Laplace transform it can be shown after some manipulations that the MGF $p_a(\gamma_1)$ of is given by [5]

$$M_a(s:\bar{\gamma}_1,\bar{\gamma}_2:m:\rho) \cong M_a(s) = \left[1 - \frac{(\bar{\gamma}_1 + \bar{\gamma}_2)}{m}s + \frac{(1-\rho)\bar{\gamma}_1\bar{\gamma}_2}{m^2}s^2\right]^{-m}$$

$$s \geq 0 \qquad (9)$$

The average error probability [14] in general is given by

$$P_e = 1 - \int_{-\pi/m}^{\pi/m} p(\theta)d\theta$$

$$where\, P(\theta) = \int_0^\infty f(\theta/R) p_o(R) dR.$$

$$P_e = \frac{\Gamma(2m+1/2)}{\varepsilon_m \sqrt{\pi}\,\Gamma(2m+1)} \cdot \frac{1}{(1-k^2)^m} \cdot \left[\frac{1}{(\gamma/m)\sin^2\left(\frac{\pi}{m}\right)}\right]^{2m}$$

[10]

$f(\theta/R)$ is the p.d.f of the detection error, $p_o(R)$ is the pdf of the envelope of the fading signal. After approximation [14]

$$P_e = \frac{\Gamma(2m+1/2)}{\varepsilon_m \sqrt{\pi}\,\Gamma(2m+1)} \cdot \frac{1}{(1-k^2)^m} \cdot \left[\frac{1}{(\gamma/m)\sin^2\left(\frac{\pi}{m}\right)}\right]^{2m} \quad (11)$$

K is the power correlation coefficient of the two set of fading from the above formulas the moment generating function is dependent upon the correlation coefficient

## 4. Simulation Results

| SIMULATION PARAMETERS | DETAILS |
|---|---|
| Order of the system (M) | 4,16 and 64 |
| Types of Modulation | QPSK, QAM |
| Data rate (Max.) | 50 Mbps |
| Coding rates | 1/2, 2/3,3/4 |
| Subcarriers for audio transmission | 52 |
| Subcarriers for video transmission | 256 |
| Number of pilot carriers | 4,8 |
| Guard interval | 800 ns |
| OFDM symbol duration | 4 µs |
| Channel bandwidth | 20MHz |

Table.1 Simulation parameters for unequal modulation

For the above mentioned parameters in a rayleigh channel the simulation has been conducted for unequal modulation orders. The resultant shows the unequal ber performance is better compare to high modulation order to all subcarriers.

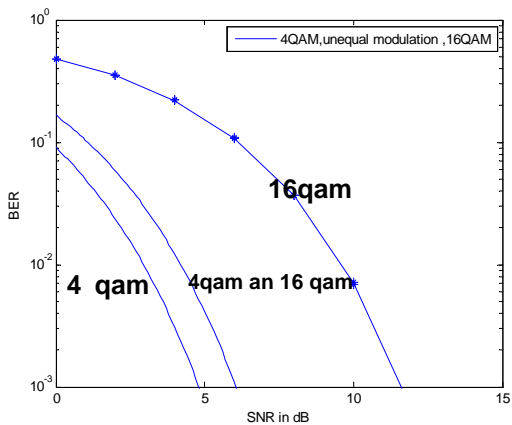

Fig. 5 Ber results for equal and unequal modulation

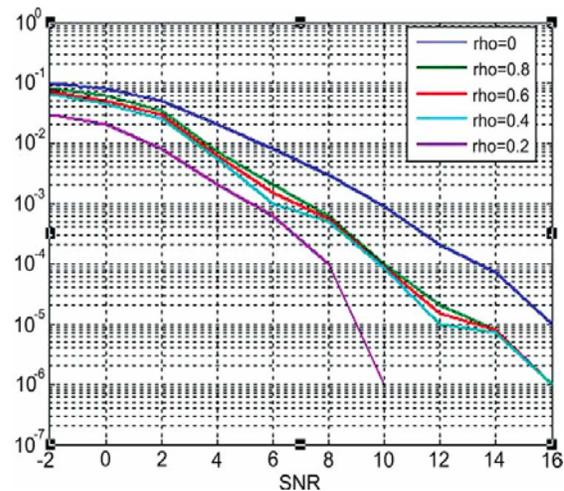

Fig. 6 Performance of SNR *vs* BER for different values of ρ

The above graph is plotted for different values of ρ for 16 qam modulation. The plot shows that as the ρ value increases the ber increases as the fading profile increases with ρ

## 5. Conclusion

The paper proposes an EESM based link adaptation algorithm in which unequal constellation orders has been used for the same sub-carrier block of an ofdm system. Simulation results have been shown for ber and fading margin with unequal constellation size. The result shows performance enhancement over equal constellation orders in ofdm system.

R. Sandanalakshmi has been teaching for 9 years in the field of electronics and communication Engineering. She is currently Asst.Professor, Department of Electronics and Communication engineering, Pondicherry engineering college. She finished her masters in the area communication engineering at pondicherry engineering college in the year 2000. She is currently pursuing her research in the field of link adaptation algorithms for multicarrier systems for next generation wireless networks.